\documentclass[aps,pra,showpacs,reprint,twocolumn,groupedaddress]{revtex4-1}

\usepackage{epsfig,amssymb,amsmath}

\usepackage{color}

\def\comment#1{}
\def\labell#1{\label{#1}}
\def\>{\rangle}\def\<{\langle}
\def\subsubsection#1{{\par\em #1:--- }}
\def\togli#1{}

\begin{document}


\title{Quantum Time} \author{Vittorio Giovannetti,$^1$ Seth
  Lloyd$^{2}$, Lorenzo Maccone$^3$}\affiliation{$^{1}$NEST-INFM \&
  Scuola Normale Superiore, Piazza dei Cavalieri 7, I-56126, Pisa,
  Italy.\\\vbox{$^{2}$RLE and Dept. of Mech. Eng., Massachusetts
    Institute of Technology,
    77  Mass. Av., Cambridge, MA 02139, USA.}\\
  \vbox{ $^3$ Dip.~Fisica ``A.  Volta'' \& INFN Sez.~Pavia,
    Universit\`a di Pavia, via Bassi 6, I-27100 Pavia, Italy.}}

\begin{abstract}
  We give a consistent quantum description of time, based on Page and
  Wootters' conditional probabilities mechanism, that overcomes the
  criticisms that were raised against similar previous proposals. In
  particular we show how the model allows to reproduce the correct
  statistics of sequential measurements performed on a system at
  different times.
\end{abstract}



\maketitle

Time in quantum mechanics appears as a classical parameter in the
Schr\"odinger equation. Physically it represents the time shown by a
``classical'' clock in the laboratory. Even though this is acceptable
for all practical purposes, it is important to be able to give a fully
quantum description of time. Many such proposals have appeared in the
literature
(e.g.~\cite{stunorm,zeht,rovelli,rovt,paw,kuchar,wilcz,ahar,sels,salecker,feynmanq}),
but none seem entirely satisfactory
\cite{kuchar,unruhwald,anderson,problemoftime,sorkk,pagereply}.  One
of these is the Page and Wootters (PaW) mechanism \cite{paw} (see also
\cite{vedral,papersinvedral,zeht,freden,cza}) which considers ``time'' as
a quantum degree of freedom by assigning to it a Hilbert space ${\cal
  H}_T$. The ``flow'' of time then consists simply in the correlation
(entanglement) between this quantum degree of freedom and the rest of
the system, a correlation present in a global, time-independent state
$|\Psi\>\>$. An internal observer will see such state as describing
normal time evolution: the familiar system state $|\psi({t})\>$ at
time $t$ arises by conditioning (via projection) the state $|\Psi\>\>$
to a time $t$ (Fig.~\ref{f:cond}), it is a conditioned state.  The PaW
mechanism was criticized in \cite{kuchar,unruhwald} and a proposal
that overcomes these criticisms \cite{gambinipullin,montevideo} used
Rovelli's evolving constants of motion \cite{rovelli,rovellibook}
parametrized by an arbitrary parameter that is then averaged over to
yield the correct propagators.  Although the end result matches the
quantum predictions \cite{esperimentotorino}, the averaging used there
amounts to a statistical averaging which is typically reserved to
unknown physical degrees of freedom rather than to parameters with no
physical significance. (A different way of averaging over time was
also presented in \cite{jonathan} to account for some fundamental
decoherence mechanism.)

Here we use a different strategy: we show that the same criticisms can
be overcome by carefully formalizing measurements through the von
Neumann prescription \cite{vonneumannbook} (which we extend to
generalized observables, POVMs).  We show how this implies that all
quantum predictions can be obtained by conditioning the global,
timeless state $|\Psi\>\>$: this procedure
gives the correct quantum propagators and the correct quantum
statistic for measurements performed at different times, features that
were absent in the original PaW mechanism \cite{kuchar,pagereply}. We
also show how the PaW mechanism can be extended to give the
time-independent Schr\"odinger equation and give a physical
interpretation of the mechanism.

What is the physical significance of the quantized time in the PaW
representation? One is free to consider the time quantum degree of
freedom either as an abstract purification space without any physical
significance or as a dynamical degree of freedom connected to some
system, or collection of systems, that represents a clock that is used
to define time. The latter point of view may describe an operational
definition of time \cite{basri,cook} that is appropriate for proper
time: it entails defining proper time as ``what is read on a clock'',
where a clock is a specific physical system (described by the Hilbert
space ${\cal H}_T$).  In what follows we do not make commitment on any
of these interpretations: our aim is only to elucidate some technical
aspects of the representation and to clarify how it can be used to
reproduce the predictions of standard quantum mechanics.

\begin{figure}[t]
\begin{center}
\epsfxsize=.9\hsize\leavevmode\epsffile{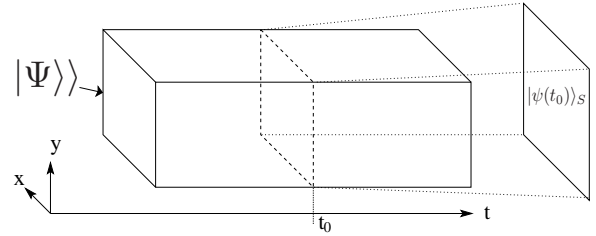}
\end{center}
\vspace{-.5cm}
\caption{Pictorial representation of the global state $|\Psi\>\>$. The
  Hilbert space of the system ${\cal H}_S$ is represented by the $x,y$
  axes, the time Hilbert space ${\cal H}_T$ by the horizontal axis.
  The state of the system $|\psi({t}_0)\>$ at time ${t}_0$ of the
  conventional formulation of quantum mechanics (dashed lines) is
  obtained by conditioning $|\Psi\>\>$ to having time ${t}_0$.}
\labell{f:cond}\end{figure}

\subsection{Review and revision of PaW}
Our proposal is an extension of PaW's mechanism
\cite{paw,w,pagereply}.  It consists in enlarging the Hilbert space
${\cal H}_S$ of the system under consideration to 
  $\mathfrak{H}:=  {\cal H}_T\otimes{\cal H}_S$ with ${\cal H}_T$
   the space of an ancillary system $T$ (we shall call it the ``clock'' system) that we assume to be
  isomorphic to the Hilbert space of a particle on a line (other
  choices are possible \cite{w,esperimentotorino}).  The latter is 
equipped with canonical coordinates $\hat{T}$ and $\hat\Omega$ with
$[\hat{T},\hat\Omega]=i$, that represent position and momentum and
(under the following restrictions) can be interpreted as the time and
energy indicator of the evolving system.  Next we
  introduce what we may call the {\it constraint} operator of the
  model, i.e.
  \begin{eqnarray} \label{GAUGE} \hat{\mathbb{J}} : =
    \hbar\hat\Omega\otimes\openone_S+\openone_{T}\otimes\hat{H}_S\;,
\end{eqnarray} 
with $\hat H_S$ the system Hamiltonian, and $\openone_S$ and $
\openone_T$ the identity operators on ${\cal H}_S$ and ${\cal H}_T$.
By construction $\hat{\mathbb{J}} $ is selfadjoint and has a
continuous spectrum that includes all possible real values as
generalized eigenvalues.  Next we identify a special set of vectors
$|\Psi\>\>$ which we call the {\it physical vectors} of the model and
which, as will be clear in the following, provides a compact, yet
static, representation of the full history of the system $S$.  They
are identified by the eigenvector equation associated with the null
eigenvalue of $\hat{\mathbb{J}} $, i.e.
\begin{eqnarray} 
\hat{\mathbb{J}}  |\Psi\>\>= 0
\labell{paw}\;,
\end{eqnarray}
where the double-ket notation reminds us that $|\Psi\>\>$ is defined
on ${\cal H}_T\otimes{\cal H}_S$.  More precisely Eq.~(\ref{paw})
defines generalized eigenvectors which (as in the case of the position
operator of a particle) are not proper elements of $\mathfrak{H}$ but
still possesses a scalar product with all the elements of such space,
inducing a representation of it.

One may interpret Eq.~\eqref{paw} as a constraint that forces the
physical vectors to be eigenstates of the total `Hamiltonian'
$\hat{\mathbb{J}} $ with null eigenvalue, consistently with a
Wheeler-DeWitt equation \cite{anderson,wdw}. Accordingly, in this model
the $|\Psi\>\>$'s are ``static'' objects which do not evolve.
The conventional state $|\psi({t})\>_S$ of the
system $S$ at time $t$ can then be obtained by conditioning a solution
$|\Psi\>\>$ of Eq.~(\ref{paw}) on having the time $t$ via projection
with the generalized eigenvectors of the time operator $\hat{T}$
(Fig.~\ref{f:cond}), i.e.
\begin{eqnarray} |\psi({t})\>_S={_{T}\<}{t}|\Psi\>\>
\labell{fullst}\;,
\end{eqnarray}
with
\begin{eqnarray} 
\hat{T} | t \rangle_T = t |t\rangle_T \;, \qquad 
{_T \langle} t' | t \rangle_T = \delta(t-t') \;.
\end{eqnarray} 
By writing \eqref{paw} in the
`position' representation in ${\cal H}_T$, 
one can easily verify that such vector indeed obeys 
Schr\"odinger equation, i.e. \cite{w,paw}
\begin{eqnarray}
_{T}\<t|\hat{\mathbb{J}}  |\Psi\>\>=0 \quad 
\Leftrightarrow \quad  i\hbar\tfrac\partial{\partial t}|\psi({t})\>_S=
\hat{H}_S |\psi({t})\>_S
\labell{sch}\;,
\end{eqnarray}
where we used the fact that $\hat{\Omega}$ is described by the differential operator. 
In a similar way we can identify the 
eigenvectors  of $\hat{H}_S$ by projecting $|\Psi\>\>$ on the (generalized) eigenstates of $\hat{\Omega}$ (i.e. the vectors 
$| \omega \rangle_T = \frac{1}{\sqrt{2\pi}} \int dt e^{i \omega t} | t\rangle_T$ with $\omega \in \mbox{Re}$). 
  Specifically, given 
\begin{eqnarray}\label{momentum} 
|\psi(\omega)\rangle_S = {_T}\langle \omega | \Psi\rangle\rangle\;,
\end{eqnarray}
with 
\begin{eqnarray} 
\hat{\Omega} | \omega \rangle_T = \omega |\omega\rangle_T \;, \qquad 
{_T \langle} \omega' | \omega \rangle_T = \delta(\omega-\omega') \;,
\end{eqnarray}
we have that 
\begin{eqnarray} \label{fffd} 
{_T}\langle \omega | \hat{\mathbb{J}}  |\Psi\rangle\rangle  =0\;\;  \Longleftrightarrow \;\; 
\hat{H}_S |\psi(\omega)\rangle_S = -\hbar \omega |\psi(\omega)\rangle_S\;,\label{sch1}
\end{eqnarray}  
which shows that the momentum representation~(\ref{momentum})  of a physical vector $|\Psi\rangle\rangle$ that solves Eq.~(\ref{paw}) 
obeys the  Schr\"{o}dinger eigenvector equation -- more precisely
for $\omega$ such that $-\hbar \omega$ equals an element of the spectrum of $H_S$, then $|\psi(\omega)\rangle_S$ is an eigenvector of $H_S$ at that eigenvalue, otherwise 
$|\psi(\omega)\rangle_S=0$.

Exploiting the fact that both $\{|t\rangle_T\}_t$ and
$\{|\omega\rangle_T\}_\omega$ provide a decomposition for the identity
operator on ${\cal H}_T$, any solution of Eq.~(\ref{paw}) can be
expressed as
\begin{eqnarray}\label{FIRST1} 
|\Psi\rangle \rangle &=& \int d t\; |t\rangle_{T} \otimes |\psi(t)\rangle_S \\
&=& \int  d \mu(\omega)\; |\omega \rangle_{T} \otimes |\psi(\omega)\rangle_S\;, 
\end{eqnarray}
with $d\mu(\omega)$ a measure on the real axis which selects those
$\omega$'s that admit a non trivial solution for  Eq.~(\ref{sch1}).  The identity~(\ref{FIRST1})
shows that the vectors $|\Psi\>\>$ provide a complete description of
the temporal evolution of the system $S$ by representing it in terms
of correlations between the latter and the degree of freedom of the
ancillary system $T$. In particular, introducing the unitary operator
$\hat{U}_S(t) = e^{-\frac{i}{\hbar} \hat{H}_S t}$ which solves
Eq.~(\ref{sch}), we get
\begin{eqnarray}\label{FIRST2} 
|\Psi\rangle \rangle &=& \int d t\; |t\rangle_{T} \otimes \hat{U}_S(t) |\psi(0)\rangle_S \\ 
&=& \hat{\mathbb{U}}  \label{FIRST3} 
 \;  |T_{ime}L_{ine}\rangle_T \otimes |\psi(0)\rangle_S 
\;, 
\end{eqnarray}
where $|\psi(0)\rangle_S$ is the state of $S$ at time $t=0$, where 
$|T_{ime}L_{ine}\rangle_T$ is the improper state of ${\cal H}_T$
obtained by superposing all vectors $|t\rangle_T$, i.e. 
 \begin{eqnarray}\label{TIME}
|T_{ime}L_{ine}\rangle_T := \int dt\;  |t\rangle_T =\sqrt{2\pi} \; |\omega=0\rangle_T \;,
\end{eqnarray}
and where $\hat{\mathbb{U}}$ is the unitary operator
\begin{eqnarray}\label{DEFU}
\hat{\mathbb{U}}&:=& \int dt |t\rangle_T\langle t| \otimes \hat{U}_S(t)\\
& = &\hat{U}_S(\hat{T}) = e^{-i  \hat{T}\otimes\hat{H}_S /\hbar}. \label{DEFU1} 
\end{eqnarray}

Before proceeding further we comment on some important technical
aspects of the PaW representation.

\subsubsection{The Zero-eigenvalue} In the construction of the PaW
model, the zero eigenvalue of $\hat{\mathbb{J}} $ seems to play a
special role in identifying the physical vectors $|\Psi\>\>$ but this
is not the case.  Indeed up to an irrelevant global phase, the
physical vectors $|\Psi\>\>$ can be identified also by imposing the
constraint
\begin{eqnarray} 
\hat{\mathbb{J}}  |\Psi\>\>= \epsilon | \Psi\>\> 
\labell{pawg}\;,
\end{eqnarray}
with real $\epsilon$. Indeed Eq.~(\ref{pawg}) can be cast in the
form~(\ref{paw}) by rigidly shifting the spectrum of $\hat{H}_S$ by $\epsilon$.

\subsubsection{Time-dependent Hamiltonians} 
%
The relevance of  Hamiltonians $\hat{H}_S(t)$
which exhibit an  explicit time dependence may be questioned at fundamental level. Still 
it is well known that the possibility of dealing with these models is extremely useful in simplifying the analysis of systems where
effective time dependent Hamiltonians arise from the interplay between local degree of freedom and an external, complex environment characterized by a semi-classical behavior (e.g. a measurement apparatus) -- see e.g.~\cite{HAMT} and references therein. 
Notably the PaW representation can be extended to incorporate also these examples by simply 
replacing the  constraint operator~(\ref{GAUGE}) with 
\begin{eqnarray} \label{GAUGE1}
\hat{\mathbb{J}}  : = \hbar\hat\Omega\otimes\openone_S+\hat{H}_{S}(\hat{T}) \;,
\end{eqnarray}  
with $\hat{H}_{S}(\hat{T})$ now an operator that acts not trivially on
both ${\cal H}_S$ and ${\cal H}_T$, obtained by formally promoting the
variable $t$ which appears in $\hat{H}_S(t)$ into the canonical
coordinate operator $\hat{T}$. Selecting the physical states
$|\Psi\>\>$ as in Eq.~(\ref{paw}) it then follows that the
decompositions~(\ref{FIRST1}), (\ref{FIRST2}), and (\ref{DEFU}) still
hold with the operator $\hat{U}_S(t,0)$ defined as
\begin{eqnarray}
\hat{U}_S(t,t_0) =\left\{ \begin{array}{ll}  \overset{\leftarrow}{\exp}[ -(i/\hbar) \int_{t_0}^t dt' \hat{H}_S(t')]  & \forall t \geq t_0\;, \\ \\
\overset{\rightarrow}{\exp}[ (i/\hbar) \int_t^{t_0} dt' \hat{H}_S(t')]  & \forall t < t_0\;, \end{array} \right.  \label{CONTROLU}
\end{eqnarray} 
where $\overset{\leftarrow}{\exp}[\int_{t_0}^t dt' ... ]$  (resp. $\overset{\rightarrow}{\exp}[ \int_{t_0}^t dt' ... ]$)  indicates the  time-ordering (resp. anti-ordering)  of the associated integral. 

\subsubsection{The initial time}  The choice of $t=0$ as the reference time in Eqs.~(\ref{FIRST2}) and (\ref{FIRST3}) is just a matter of convention. Indeed an equivalent way to express $|\Psi\>\>$ is the following
\begin{eqnarray}\label{FIRST2new} 
|\Psi\rangle \rangle &=& \int d t\; |t\rangle_{T} \otimes \hat{U}_S(t,t_0) |\psi(t_0)\rangle_S \\ 
&=& \hat{\mathbb{U}}_{t_0}  \label{FIRST3new} 
 \;  |T_{ime}L_{ine}\rangle_T \otimes |\psi(t_0)\rangle_S 
\;, 
\end{eqnarray}
where now $|\psi(t_0)\>$ is the state of the system at time $t_0$ and where
\begin{eqnarray}\label{DEFUnew}
\hat{\mathbb{U}}_{t_0} &:=& \int dt |t\rangle_T\langle t| \otimes \hat{U}_S(t,t_0) = \hat{\mathbb{U}}\; [ \openone_T \otimes  \hat{U}_S(t_0,0) ] \;,\nonumber \\
\end{eqnarray}
(the identity is valid also in the case of time-dependent Hamiltonian
$\hat{H}_S(t)$).
\\

\subsubsection{Propagators} As anticipated, the physical vectors
$|\Psi\>\>$ give a compact description of the system dynamical
evolution in terms of a superposition of components each associated
with a different time measured by an external clock described by
${\cal H}_T$. In particular, suppose we want to calculate the
propagator between a state $|I\>_S$ at time $t_I$ and a state $|F\>_S$
at time $t_F$, i.e. the quantity ${\cal G}(F,t_F; I,
t_I):={}_S\<F|\hat{U}_S(t_F,t_I)|I\>_S$.  In the PaW formalism this
can be obtained by simply identifying $t_0$ with the time $t_I$ and
$|\psi(t_0)\>_S$ with $|I\>_S$ in Eq.~(\ref{FIRST2new}) (this fixes
the initial condition of the system trajectory) and then projecting
the associated $|\Psi\>\>$ on $|t_F\>_T\otimes |F\>_S$, i.e.
\begin{eqnarray} 
 {\cal G}(F,t_F; I, t_I) = ( {_T \<} t_F| \otimes {_S \<} F| ) | \Psi\>\> \;. 
\end{eqnarray} 
One of the criticisms to the PaW mechanism is the fact that it did not
seem to be able to reproduce the correct propagators \cite{kuchar}. He
we have shown how the correct propagators can emerge.

\subsubsection{About conditioning}
In the PaW representation the physical vectors identified by
Eq.~(\ref{paw}) ideally should describe a joint state of $S$ and of
the clock system $T$. Accordingly, given $\{ |a\>_S\}$ a complete
orthonormal basis for $S$, the quantities $({_T\< t|} \otimes
{_S\<}a|)|\Psi\>\>$ should correspond to proper amplitude
joint-probability distributions associated with the probability of
finding $|a\>_S$ on $S$ {\em and} $|t\>_T$ on $T$.  In this framework
it makes sense to interpret the vector (\ref{fullst}) as the
conditioned state of $S$ obtained by forcing $T$ to be on $|t\>_T$.
Similarly one would like to interpret ${_S\<}a|\Psi\>\>$ as the state
of the clock conditioned by forcing $S$ to be on $|a\>_S$.  This last
assumption however is problematic because, being $|\Psi\>\>$ an
improper element of $\mathfrak{H}$, the vector ${_S\<}a|\Psi\>\>$ as
well as the function $|({_T\< t|} \otimes {_S\<}a|)|\Psi\>\>|^2$, do
not admit a proper normalization, forcing us to assign a uniform
distribution to the time variable $t$ -- no other choice being allowed
by the representation.  One can fix this by replacing
Eq.~(\ref{FIRST2}) with a normalized element of $\mathfrak{H}$, i.e.
with vectors of the form
\begin{eqnarray}
  |\Phi\>\>=\int \label{FGFG}
  dt\:\phi(t)\;  |t\>_T \otimes |\psi(t)\>_S\;,
\end{eqnarray}
with $|\psi(t)\>_S$ a normalized vector of ${\cal H}_S$ (i.e. $\|
|\psi(t)\>_S\|=1$) and with $\phi(t)$ a square integrable function
that guarantees the normalization condition for $|\Phi\>\>$ , i.e.
$\| |\Phi\>\>\|^2= \int dt |\phi(t)|^2=1$. Note that $|\Phi\>\>$ of Eq.~\eqref{FGFG} is the most
general state of $\mathfrak{H}={\cal H}_S\otimes{\cal H}_T$. Imposing
next $|\psi(t)\>_S$ to describe the evolution of the initial state
$|\psi(0)\>_S$ under the action of the system Hamiltonian $\hat{H}_S$
we can then write
\begin{eqnarray}
  |\Phi\>\>=\hat{\mathbb{U}}  \label{FIRST333} 
  \;  |\phi \rangle_T \otimes |\psi(0)\rangle_S\;, 
\end{eqnarray}
which replaces Eq.~(\ref{FIRST3}) by substituting the improper vector
$|T_{ime}L_{ine}\rangle_T$ with the properly normalized state
$|\phi\>_T:=\int dt\,\phi({t})|{t}\>_{T}$ -- the operator
$\hat{\mathbb{U}}$ is still defined as in Eq.~(\ref{DEFU}).  For any
assigned $\phi(t)$, the identity~(\ref{fullst}) is now replaced by
\begin{eqnarray}
|\psi({t})\>_S= {}_{T}\!\<{t}|\Phi\>\>/\phi({t})
\labell{qbayes}\;
\end{eqnarray}
which, by noticing that $\phi(t)$ is the amplitude probability of
finding the clock on $|t\>_S$ when measuring $|\Phi\>\>$, makes
explicit the conditioning nature of $|\psi(t)\>_S$: such vector is
obtained as an application of a Bayes rule for probability amplitudes,
where the numerator gives the joint statistics of measurement on $S$
and on $t$ and the denominator describes the statistics of measurement
on $t$ only. At variance with the state of Eq.~(\ref{FIRST2}), the
representation (\ref{FIRST333}) finally allows for a proper definition
of the state of the clock: the reduced density matrix
Tr$_S[|\Phi\>\>\<\<\Phi|]$ is now well behaved. These considerations
imply that $\phi({t})$ is the weight that represents the probability
amplitude that the system is found at time $t$, namely, in a sense, it
is the probability amplitude that the system ``exists'' at such time.
[Clearly, a suitable regularization is implicit in the expression
\eqref{qbayes}, to avoid Borel-Kolmogorov type paradoxes that arise
when one conditions on something that has null probability.]

In view of the above results, we can interpret the representation
$|\Phi\>\>$ of~(\ref{FIRST333}) as a regularized version of the
original PaW representation $|\Psi\>\>$ of \eqref{FIRST1}, since it
satisfies the normalization $\<\<\Phi|\Phi\>\>=1$ on the joint system,
which is the St\"uckelberg normalization \cite{stunorm}. In fact,
following the conventional technique used for regularizing the
eigenstates of operators with continuous spectrum, the PaW state
\eqref{FIRST1} can be replaced, for example, by a normalized state
\eqref{FGFG} with Gaussian weight
$\phi({t})\equiv\phi_n({t})=(2/n\pi)^{1/4}\exp(-{t}^2/n)$. Then, using the
Weyl criterion \cite{weyl}, one can conclude that $\lambda=0$ is an
essential eigenvalue of the self-adjoint operator $\hat{\mathbb{J}}$,
since \begin{eqnarray} \| (\hat{\mathbb{J}}-\lambda)|\Phi_n\>\>\|\to
  0\mbox{ for }n\to\infty
\labell{l}\;
\end{eqnarray}
where $|\Phi_n\>\>=\int dt\:\phi_n(t)|{t}\>_T|\psi({t})\>_S$ is
\cite{ess} a Weyl sequence, i.e.~a normalized sequence of Hilbert
space vectors that converges weakly to 0, namely
$\<\<\theta|\Phi_n\>\>\to 0$ for $n\to\infty$ for all
$|\theta\>\>\in{\cal H}_{T}\otimes{\cal H}_S$.  Moreover, the
un-normalized PaW state $|\Psi\>\>$ is obtained for $n\to\infty$ as
$(n\pi/2)^{1/4}|\Phi_n\>\>\to|\Psi\>\>$.

The representation \eqref{FIRST333} allows for a constraint
description analogous to~(\ref{paw}) obtained by adding $\hat{H}_S$ to
the non-Hermitian correction term
${i\dot\phi(\hat{T})}/{\phi(\hat{T})}$ (the dot representing time
derivation) yielding
\begin{eqnarray}
&&\Big( (\hbar\hat\Omega+i\hbar
\frac{\dot\phi(\hat{T})}{\phi(\hat{T})})\otimes\openone_S
+\openone_{T}\otimes\hat{H}_S\Big)|\Phi\>\>=0
\labell{lal1}\;,\mbox{ i.e.}\\&&\nonumber
\Big((\phi(\hat {T})\hbar\hat\Omega+[\phi(\hat
{T}),\hbar\Omega])\otimes\openone_S+
\phi(\hat {T})\otimes\hat H_S\Big)|\Phi\>\>=0. \nonumber 
\end{eqnarray}
Alternatively, one can still retain the constraint equation
\eqref{paw} if one supposes that the Schr\"odinger equation applies
also to non-normalized states $|\psi'({t})\>=\phi({t})|\psi({t})\>$ as
$\hat H_S|\psi'({t})\>=i(\partial/\partial{t})|\psi'({t})\>$, in
analogy to the action of the momentum operator on non-normalized
wavefunctions (such as the components of spinors). Both of these
approaches are extensions of conventional quantum mechanics, which
deals only with states that are normalized at all times.

We stress that, while working on this theoretical framework may be
have some appeal, the approach is not fully satisfactory as for
instance the choice of $\phi(t)$ is completely arbitrary and there is
no indication in the conventional theory on how to fix it. The fact
that $\phi(t)$ is non-unique is a consequence of the freedom that one
has in quantum mechanics to choose any vector of the Hilbert space as
representing a valid state of the system, as long as it does not
violate physical or dynamical constraints.

\subsubsection{Physical interpretation}
We briefly comment here on the physical interpretation of the
additional Hilbert space ${\cal H}_T$. One can interpret it as an
abstract `purification' space without physical relevance. However, an
operational definition of proper time \cite{basri} as ``such that is
measured by a clock'' requires some physical system that acts as a
clock. In contrast to the conventional formulation of quantum theory,
the above formalism naturally accommodates it: ${\cal H}_T$ is the
Hilbert space of such system. Clearly the particular form of ${\cal
  H}_T$ employed above is an idealization where the clock is
isomorphic to a particle on a line \cite{salecker}. Other choices
\cite{w,esperimentotorino} are a straightforward modification of the
above theory. This approach is consistent with a relational point of
view, where the only physically relevant quantities are events defined
as coincidences in spacetime \cite{gravitation} such as the
correlations between observables and what is shown on a local clock
(e.g.~\cite{rovellibook}, Sec.~2.3).

Is the above physical definition of proper time sufficient to identify
time, i.e.~coordinate time? It is for Newtonian mechanics (coordinate
time = proper time) and for special relativity (coordinate time =
proper time of a static inertial observer). In general relativity any
observer can identify the coordinate time from its own proper time if
the metric is known and considered as a classical degree of freedom
\cite{cook,klioner}, even though the coordinate time has no physical
meaning \cite{gravitation} and it is impossible to synchronize local
clocks meaningfully (i.e.~so that two clocks synchronized to a master
clock are synchronized among themselves) \cite{basri}. The case in
which the metric is considered as a quantum degree of freedom is
currently an open problem and clearly beyond the scope of the present
work.

When one considers time as a dynamical variable, an apparent
contradiction arises (\cite{peresbook}, sec.~8-6): if one interprets
momentum as the generator of space translations and the Hamiltonian as
the generator of time translations, then one would expect that the
Hamiltonian always commutes with the momentum, since these two
translations are independent. Why is this untrue in general? In the
conventional formalism, time is not a dynamical variable, so the
unitary transformations generated by the Hamiltonian are not
symmetries of the system. In contrast, in the PaW formalism, time {\em
  is} a dynamical variable, but the generator of its translations is
$\hat\Omega$, not the system Hamiltonian $\hat H_S$, and $\hat\Omega$
indeed commutes with the system momentum (it acts on a different
Hilbert space). The above apparent contradiction is thus resolved in a
different manner.

\subsection{Measurements}

At variance with what is typically believed
(e.g.~\cite{kuchar,pagereply}), the PaW formalism, appears to be
particularly well suited to describe in a compact form the statistics
of measurements which are performed sequentially on a system of
interest.

To show this explicitly let us first analyze the case where a
measurement is performed at time $t_1$ on the system $Q$.  We begin
adopting the von Neumann formulation of a measurement apparatus
\cite{vonneumannbook}, describing the process in terms of a memory
system $M$ that is in a fiducial state ``ready'' $|{r}\>_M$ before the
measurement and which will be in a state $|{a}\>_M$ that contains the
measurement outcome after. In other words we describe the measurement
as an instantaneous transformations which at time $t_1$ induces the
following unitary mapping:
\begin{eqnarray}
|\psi(t_1)\>_Q\otimes | {r}\>_M\longrightarrow
\sum_a \hat{K}_a |\psi(t_1)\>_Q \otimes |{a}\>_M\;, \labell{measurement}
\end{eqnarray}
where $\{ \hat{K}_a\}$ are Kraus operators fulfilling the
normalization condition $\sum_a \hat{K}_a^\dag \hat{K}_a =
\hat{\openone}$. Projective nondegenerate von Neumann measurements are
the special case in which $\hat K_a=|a\>\<a|$ are projectors on the
eigenspaces relative to the eigenstates $|a\>$ of the observable. In
this specific case, Eq.~\eqref{measurement} becomes
\cite{vonneumannbook}
\begin{eqnarray}
|\psi(t_1)\>_Q\otimes | {r}\>_M\longrightarrow
\sum_a \psi_a({t}_1) |a\>_Q \otimes |{a}\>_M\;, 
\labell{measurement1}\;
\end{eqnarray}
with $ \psi_a({t}_1):=\<a|\psi({t}_1)\>$. Accordingly, the
probability of getting the outcome $a$ is given by
\begin{eqnarray} \label{defip} P(a|t_1) := \| \hat{K}_a
  |\psi(t_1)\>_Q\|^2
  \;, 
\end{eqnarray} 
with $\| |v\>\| =\sqrt{\< v| v\>}$ the norm of the vector $|v\>$, and
\begin{eqnarray} \label{defphi} | \phi_a \rangle_Q := \hat{K}_a
  |\psi(t_1)\>_Q/\sqrt{P(a|t_1)} \;,
\end{eqnarray} 
is the vector which describes the state of the system $Q$ immediately
after such event has been recorded by the memory $M$. \togli{Some redundancy
in the memory degrees of freedom is necessary to identify the
measurement basis, through einselection \cite{darwinism}. }In the
general setting, Eq.~(\ref{measurement}) defines the statistical
properties of a Positive Operator Valued Measure (POVM), see
e.g.~\cite{HOLEVOBOOK}.


The process described above can now be cast in the PaW formalism by
redefining $S$ to include both the system to be measured $Q$ and the
ancillary memory system $M$. In this context we shall assume no
interactions between $Q$ and $M$ apart from a strong (impulsive)
coupling between $Q$ and $M$ at time $t_1$ that is responsible for the
mapping~(\ref{measurement}). Adopting the time-dependent
description~(\ref{CONTROLU}) we write 
\begin{eqnarray}\label{HAMMIS}
\hat{H}_S(t) = \hat{H}_Q(t)   + \delta(t-t_1) \hat{h}_{QM} \;,
\end{eqnarray} 
where $\hat{H}_Q(t)$ is the (possibly time-depedent) free Hamiltonian
of $Q$, where $\delta(x)$ is the Dirac delta function, while
$\hat{h}_{QM}$ is related to the unitary $\hat{V}_{QM}$ responsible of
the mapping Eq.~(\ref{measurement}) via the identity $\hat{V}_{QM} :=
e^{-\frac{i}{\hbar} \hat{h}_{QM}}$ (since $M$ is a memory we assume no
free dynamics for it).  With this choice
\begin{eqnarray}
\hat{U}_S(t,t_0) =\left\{ \begin{array}{ll} \hat{U}_Q(t,t_0)  & \forall t < t_1\;, \\ \\
\hat{U}_Q(t,t_1) \hat{V}_{QM} \hat{U}_Q(t_1,t_0)  & \forall t > t_1\;, \end{array} \right.
\;,  \label{CONTROLU111}
\end{eqnarray} 
where $\hat{U}_Q(t,t')$ is the operator which gives the free evolution
of $Q$ defined as in Eq.~(\ref{CONTROLU}) through the Hamiltonian
$\hat{H}_Q(t)$~\cite{NOTA}. Accordingly, Eq.~(\ref{FIRST2new}) becomes
\begin{eqnarray}
  &&
  |\Psi\>\>=\int_{-\infty}^{t_1}\!
  dt\;|t\>_T\otimes  |\psi(t)\>_Q\otimes |r\>_M \nonumber\\&& 
 + \int_{t_1}^{\infty}\!
  dt\;|t\>_T \otimes \sum_a \hat{U}_Q(t,t_1) \hat{K}_a |\psi (t_1)\>_Q\otimes
|a\>_M \labell{pur}\;,
\end{eqnarray}
where  for $t< t_1$, $|\psi(t)\rangle_Q= \hat{U}_Q(t,t_0) | \psi(t_0)\rangle_Q$ is the state of $Q$ at time $t$ prior of the  measurement stage.
In this framework the probability that, at a given time $t$ measured by the ancillary system $T$,  a certain outcome $a$ will be  registered by the memory $M$  can be formally expressed as~\cite{NOTA}
\begin{eqnarray}
P(a| t) &=& \|( {_T \<} t| \otimes {_M \<} a | ) |\Psi\>\> \|^2 \label{DEFPA}.
 \end{eqnarray}  
 As a consequence of the impulsive coupling we have assumed in
 describing the measurement process, Eq.~(\ref{DEFPA}) is a step
 function which exhibits a sharp transition at the measurement time
 $t=t_1$: for smaller values of $t$, the probability of getting a
 certain outcome $a$ on $M$ does not depend upon $Q$ yielding $P(a| t)
 = |{_M \<} a| r\>_M|^2 $, the resulting statistics being only
 associated with the the ready state of the memory; for ${t}> t_1$
 instead, $P(a| t)$ coincides with the value (\ref{defip}): it only
 depends upon the statistical uncertainty of the state of the system
 $Q$ at time $t_1$ and it remains constant in time due to the fact
 that we have explicitly suppressed any dynamical evolution on $M$.
 
The above framework immediately
extends to the case where different measurements are performed at different times, giving the correct
transition probabilities. This was lacking \cite{kuchar} in the PaW
proposal. In fact, the global state of a system where a measurement of
$A$ at time ${t}_1$ and of $B$ at a later  time ${t}_2>t_1$  is performed can be expressed within the formalism  by adding an extra memory element $M'$ which stores the information associated with the second measurement. 
Accordingly, we replace Eq.~(\ref{HAMMIS}) 
with 
\begin{eqnarray}
\hat{H}_S(t) = \hat{H}_Q(t)   + \delta(t-t_1) \hat{h}_{QM} + \delta(t-t_2) \hat{h}_{QM'} \;,
\end{eqnarray} 
with $\hat{h}_{QM'}$ responsible for the unitary coupling
$\hat{V}_{QM'}$ associated with the measurement of $B$.  With this
choice for all $t>t_2$, Eq.~(\ref{CONTROLU111}) gets replaced by
$\hat{U}_Q(t,t_2) \hat{V}_{QM'} \hat{U}_Q(t_2,t_1) \hat{V}_{QM}
\hat{U}_Q(t_1,t_0)$ while the state $|\Psi\>\>$ becomes
\begin{eqnarray}
&&|\Psi\>\>=\int_{-\infty}
^{{t}_1}+\int_{{t}_1}^{{t}_2}+
\int_{{t}_2}^\infty
\! dt\:|t\>_T \labell{double}
\otimes\\ \nonumber&&  \sum_{ab}\hat{U}_Q(t,t_2) \hat{K}_b
\hat{U}_Q(t_2,t_1) \hat{K}_a |\psi (t_1)\>_Q
\otimes |a\>_M\otimes |b\>_{M'}\;,
\end{eqnarray}
where the first two integrals have the same integrands as the
left-hand-side of \eqref{pur}, $M'$ is the memory where the $B$
outcome is stored, $b$ and $\hat{K}_b$ the corresponding outcomes and
Kraus operators. It is worth observing that the probability $P(a|t)$
of getting an outcome $a$ at time $t$ is not affected by the presence
of the second measurement: this quantity can still be computed by
projecting $|\Psi\>\>$ on $|t\>_T \otimes |a\>_M$ and assumes the same
value given in Eq.~(\ref{DEFPA}).  Similarly the joint probability
that at time $t$ the two memories will record a certain outcome $a$
and $b$, respectively, can be computed as
\begin{eqnarray} 
P(b, a|t) =  \|( {_T \<} t| \otimes {_M \<} a | \otimes {_{M'} \< } b|  ) |\Psi\>\> \|^2  \label{DEFBA}\;.
\end{eqnarray}
As in the case of Eq.~(\ref{DEFPA}) 
also this is a step function. 
In particular for $t \geq t_2$ it assumes the value \begin{eqnarray}\label{DEFBAT1}
P(b,a|t) &=& \|  \hat{K}_b
\hat{U}_Q(t_2,t_1) \hat{K}_a |\psi (t_1)\>_Q\|^2 \nonumber \\
&=&  \|  \hat{K}_b |\phi_a (t_2,t_1)\>_Q\|^2 \;  \|  \hat{K}_a |\psi (t_1)\>_Q\|^2
\;, 
\end{eqnarray} 
where in the second line we used Eqs.~(\ref{defip}) and (\ref{defphi})
and where $ |\phi_a (t_2,t_1)\>_Q = \hat{U}_Q(t_2,t_1)|
\phi_a\rangle_Q$ is the evolved via $H_Q(t)$ of the state
$|\phi_a\rangle_Q \propto \hat K_a|\psi({t}_1)\>_Q$ of the system $Q$
when the first measurement yields the outcome $a$.  The quantity $ \|
\hat{K}_b |\phi_a (t_2,t_1)\>_Q\|^2$ is nothing but that the
conditional probability $P\big[(b|t) {\big|} (a| t_1)\big]$ of getting
the outcome $b$ when measuring $B$ on $Q$ given that the outcome $a$
was registered by the first measurement performed at time $t_1$.
Invoking Eq.~(\ref{DEFPA}) we notice that it obeys the identity 
\begin{eqnarray}
    \labell{transition}P\big[(b|t) {\big|} (a| t_1)\big]=
    \frac{P( b,{ a}| t)}{P({a}|t_1)}\;. \label{BAYES}
\end{eqnarray}
This allows us to  identify $P(b,{ a}|t)$ with the joint probability
$P\big[(b|t), (a|t_1)\big]$ of getting $b$ on $M'$ at time $t$ and $a$
on $M$ at time $t_1$. In fact we have 
\begin{eqnarray}
P\big[(b|t), (a|t_1)\big] =  P\big[(b|t) {\big|} (a| t_1)\big] P({a}|t_1) =P( b,{ a}|t)\;, \nonumber\\\label{COND}
\end{eqnarray} 
where in writing the first identity we used the Bayes rule.
It is  worth stressing that Eq.~(\ref{DEFBA}) can also be applied for times $t$ prior than $t_2$. In this case we get 
$P(b,a|t) = P(a|t) \;   |{_{M'} \<} b| r\>_{M'}|^2$ with $P(a|t)$ as in~(\ref{DEFPA})   and with  $ |{_{M'} \<} b| r\>_{M'}|^2$  accounting for the statistical distribution of the ready state of $M'$. 
Similarly we can extend  Eq.~(\ref{COND}) for $t\in]t_2,t_1]$ -- indeed  one can easily verify that in this case $P[(b|t),(a|t_1)]=P( b,{ a}|t)=P( b,{ a}| t_1)$.

From the above expressions we can finally compute the probability
$P(b|t)$ of getting an outcome $b$ at time $t\geq t_2$, irrespective
of the outcome of the $A$ measurement.  This is given by the marginal
distribution obtained by tracing $P(b, a|t)$ with respect to the $a$
variable, i.e.
\begin{eqnarray}
P(b|t) = \sum_a P(b, a|t)  = 
  \|( {_T \<} t|  \otimes {_{M'} \< } b|  ) |\Psi\>\> \|^2\;,
 \label{PB}
\end{eqnarray} 
where the second identity 
 is a consequence of the fact that $\{ |a\rangle_M\}$ is a complete set for $M$. 
 For $t<t_2$ (i.e. prior then the measurement event $B$) this is just $P(b|t) = |{_{M'} \<} b| r\>_{M'}|^2$, while 
  for ${t}> t_2$ (see~ Eq.~(\ref{DEFBAT1})) we get 
 \begin{eqnarray}
 P(b|t) = \sum_a   \|  \hat{K}_b \hat{U}_Q (t_2,t_1)  \hat{K}_a |\psi (t_1)\>_Q\|^2 \;.
 \end{eqnarray} 
Equations~(\ref{DEFPA}), (\ref{DEFBA}) and (\ref{PB}) are the main results of this section and are summarized in Table~\ref{table}.
\begin{table}[t] 
\begin{tabular}  {|c|} \hline\\ 
Joint probability of getting $b$ and $a$ at time $t$: \\
$P(b, a|t) =  \|( {_T \<} t| \otimes {_M \<} a | \otimes {_{M'} \< } b|  ) |\Psi\>\> \|^2$ \\ \\ \hline \\ 
Probability of getting  $a$ at time $t$: \\
$P(a|t) = \sum_b P(b, a|t)=  \|( {_T \<} t| \otimes {_M \<} a | ) |\Psi\>\> \|^2$ 
\\ \\
Probability of getting  $b$ at time $t$: \\
$P(b|t)=  \sum_a P(b, a|t)=  \|( {_T \<} t| \otimes {_{M'} \<} b | ) |\Psi\>\> \|^2$ 
\\ \\ \hline \\
Joint prob. of getting $b$ at time $t''$ and $a$ at time $t'(<t'')$: \\
$P[(b| t''), (a|t')] = P(b, a| t'') = \|( {_T \<} t''| \otimes {_M \<} a | \otimes {_{M'} \< } b|  ) |\Psi\>\> \|^2$ \\ \\ 
Cond. prob. of getting $b$ at time $t''$ given $a$ at time $t'(<t'')$: \\
$P[(b| t'')|(a| t')] = \frac{P(b, a|t'')}{P(a|t')} =\frac{ \|( {_T \<} t''| \otimes {_M \<} a | \otimes {_{M'} \< } b|  ) |\Psi\>\> \|^2}{\|( {_T \<} t'| \otimes {_M \<} a | ) |\Psi\>\> \|^2}$ \\ \\ 
\hline
 \end{tabular} 
\caption{Probability distribution associated with a two measurement events. The first thee identities hold irrespectively from the ordering of the events (i.e. first $A$ and then 
$B$, or first $B$ and then $A$). The last two instead assume a specific ordering, i.e. first $A$ and then $B$. \label{table}} 
\end{table}

More generally, consider the case where $Q$ undergoes to a sequence of
measurements $A_1$, $A_2$, $\cdots$, $A_N$ performed at times $t_1$,
$t_2$, $\cdots$, $t_N$ which, for convenience, we can assume to be
ordered so that $t_{n+1} > t_n$ for all $n=1, 2, \cdots, N$.  We
describe this by adding $N$ memory systems $M_1$, $M_2$, $\cdots$,
$M_N$, each initialized into a ready state $|r\>_{M_n}$ and which
couple with $Q$ though the time-dependent Hamiltonian
\begin{eqnarray}\label{HMAN}
\hat{H}_S(t) = \hat{H}_Q(t)   +\sum_{n=1}^N  \delta(t-t_n) \hat{h}_{QM_n} \;.
\end{eqnarray} 
Via Eqs.~(\ref{CONTROLU}) and (\ref{FIRST2new}) this defines the physical vector $|\Psi\>\>$ of the problem. 
In this context the joint probability that at time $t$ the memory will register  a certain string $\vec{a}:=(a_{1}$, $a_{2}$, $\cdots$, $a_{N})$  of outcomes can then be computed as
\begin{eqnarray} 
P(\vec{a}|t) &=&  
\|( {_T \<} t|\otimes {\<}\vec{a}  |) \; |\Psi\>\> \|^2 
\label{DEFBABBBnn} \;,\end{eqnarray} 
with  $|\vec{a}\> := |a_{1} \>_{M_{1}} \otimes |a_{2} \>_{M_{2}} \otimes \cdots \otimes  |a_{N} \>_{M_{N}}$. 
Exploiting the Bayes rule argument one can also observe that, given a collection of times $t_1'< t_2'< \cdots <t_N'$, Eq.~(\ref{DEFBABBBnn}) provides the  joint probability associated with the events of obtaining the outcome $a_n$ at time $t_n'$, i.e.  
\begin{eqnarray} 
P\big[(a_1|t_1'), (a_2|t_2'), \cdots , (a_N| t_N')\big] &=&  P(\vec{a}|t'_N).
\label{1DEFBABBBnn}\end{eqnarray} 
Similarly given a subset $M_{j_1}$, $M_{j_2}$, $\cdots$, $M_{j_K}$ formed by $K\leq N$ different memories, the joint probability $P(\vec{a}^{(K)}|t)$  that at time $t$ they will record 
certain events $\vec{a}^{(K)}:=(a_{j_1}$, $a_{j_2}$, $\cdots$, $a_{j_K})$ is obtained by considering the associated marginal of~(\ref{DEFBABBBnn}), i.e. 
\begin{eqnarray} 
P(\vec{a}^{(K)}|t) &=&   \sum  P(\vec{a}|t) = 
\|( {_T \<} t|\otimes {\<}\vec{a}^{(K)}  |) \; |\Psi\>\> \|^2 \;, \nonumber \\ 
\label{DEFBABBB} \end{eqnarray} 
where in the first identity the sum is performed over all components of $\vec{a}$ which are not involved in the definition of $\vec{a}^{(K)}$ and where
 $|\vec{a}^{(K)}\> := |a_{j_1} \>_{M_{j_1}} \otimes |a_{j_2} \>_{M_{j_2}} \otimes \cdots \otimes  |a_{j_K} \>_{M_{j_K}}$. 

\subsection{Overcoming criticisms}
Here we give an overview of the main criticisms to the PaW mechanism
and to the conditional probability interpretation and show how our
proposal overcomes them.

There are two main criticisms to the PaW mechanism
\cite{kuchar,unruhwald,gambinipullin}. The first refers to
superselection \cite{aharonovsuskind,rudolphrmp,wiseman}: the
observables of a theory must commute with the theory's constraints.
Whenever one of the constraints is the total energy, such as in
canonical general relativity, then all observables must be stationary
as they commute with the Hamiltonian. In the Schr\"odinger picture
this translates into static physical states, which contrasts with
obvious experimental evidence and is the root of the problem of time
\cite{anderson,problemoftime,kuchar}. The second refers to the fact
that the PaW mechanism is not able to provide the correct propagators,
or the correct two-time correlations \cite{kuchar}: after the first
time measurement, the clock remains ``stuck''.  We have already shown
how these criticisms can be overcome: the first is solved by using a
global state $|\Psi\>\>$ that is independent of time and observing
that internal observers will use conditioned states, the second by
using conditioning through a von Neumann description of the
measurement interaction.  In a sense, our prescription fulfills Page's
desiderata \cite{pagereply} in showing that the second objection can
be overcome by interpreting a measurement at two different times (or,
equivalently a preparation followed by a measurement) as a single
measurement that acts both on the system and on the degrees of freedom
that store the earlier measurement outcome.  It is a sort of
purification of the time measurements and implements Wheeler's
operationalist stance that ``the past has no existence except as it is
recorded in the present''.  \cite{wheelerpast,pagereply}.

Further criticisms were proposed in \cite{unruhwald}, where it was
noted that {\em (i)}~interpretive problems cannot be alleviated
incorporating observers into the theory; {\em (ii)}~in a constrained
theory where one of the constraints is the energy (such as the
Hamiltonian formulation of General Relativity), all observables
commute with the Hamiltonian and no time dependence is possible. This
is true also for two-time correlation functions and propagators
\cite{kuchar}; {\em (iii)}~no dynamical variable can correlate
monotonically with `Heraclitian' time if the Hamiltonian is lower
bounded; {\em (iv)} only time is appropriate for conditioning the
state: for example, space may be inappropriate for setting the
conditions since a system may occupy the same position multiple times
or never.

Our mechanism replies to {\em (ii)} by indeed carefully incorporating
the observers into the theory, thereby overturning {\em (i)}. In fact,
there are two points of view: the external observer (clearly, a
hypothetical entity whenever the whole universe is considered) and the
internal observer. The Hamiltonian constraint refers to the external
observer's point of view, who sees the whole laboratory (or universe)
as a static system whose state is an eigenstate of its global
Hamiltonian. That, however, does not prevent the internal observer
from observing evolving systems, time-dependent measurement outcomes
and Born-rule induced wave-function collapses, as shown above. In a
sense, the ``relativity'' philosophy is extended also to quantum
mechanics: states and measurements are relative to the observer
\cite{everettrmp,zeh}, just as time and space are relative. Indeed, we
showed above how internal observers recover the correct two-time
correlations and propagators. As regards to objection {\em (iii)},
indeed if we want to describe a non-periodic time variable that takes
all values (the `Heraclitian' time), we must use an unbounded
`Hamiltonian': if one considers Eq.~\eqref{paw} as a sort of
Wheeler-De Witt equation, that Hamiltonian is unbounded (it contains a
`momentum' operator $\hat\Omega$). We remark that other choices may
lead to `periodic time' coordinates, but that is acceptable in
specific cosmologies: it is certainly not surprising that a system
with finite global energy will have periodic evolution.  In these
cases, except as an approximation internal observers will not be able
to use a Schr\"odinger equation, as predicted in \cite{unruhwald}.
They must employ a more general dynamical equation. As regards to
point {\em (iv)}, time's role in the conditioning to achieve
conventional quantum mechanics is made transparent by our formulation,
which can be used to show its identical role to space regarding
conditioning. In fact, just as for space, it is possible that a system
never occupies a given time, or that it occupies the same time at two
different locations if it follows a closed timelike curve, whose
existence is predicted by general relativity \cite{goedel} and studied
also in the context of quantum mechanics \cite{ctc}. So, while indeed
only time is the appropriate quantity for conditioning to obtain the
conventional theory, quantizing time with our mechanism is a viable
pathway to the unconditioned theory.

\subsection{Conclusions}
Here we modified the PaW mechanism to give a quantization of time, and
showed how the conventional quantum mechanics and the correct quantum
predictions (e.g.~regarding propagators and measurement statistics)
arise from a quantum Bayes rule by conditioning the global state
$|\Psi\>\>$ to a specific time. We emphasize that our approach can
quantize time for completely arbitrary quantum systems
$|\psi({t})\>_S$.  As such, we can also provide a description of
quantum field theory with a quantum clock.

We acknowledge funding from the FQXi's grant ``The Physics of What
Happens''.

LM acknowledges useful discussions with B. Bertotti, J. Pullin and H.
Nikolic.

\end{document}